\documentclass[superscriptaddress,nobibnotes,amsmath,amssymb,notitlepage,twocolumn,prb,reprint,longbibliography]{revtex4-1}

\usepackage{setspace} 
\usepackage{nicefrac} 
\usepackage{xfrac}    
\usepackage{graphicx} \graphicspath{{./img/}} 
\usepackage{amsmath}
\usepackage{color}
\usepackage{amsmath}
\usepackage{amssymb}
\usepackage{verbatim}
\usepackage{latexsym}
\usepackage{enumerate} 
\usepackage{bm} 
\usepackage{ulem} 

\begin{document}

\title{
Designing Antiferromagnetic Spin-1/2 Chains in Janus Fullerene Nanoribbons 
}



\newcommand{\TCM}{Theory of Condensed Matter Group, Cavendish Laboratory, University of Cambridge, J.\,J.\,Thomson Avenue, Cambridge CB3 0HE, UK}
\newcommand{\HarvardFAS}{Harvard University, Cambridge, Massachusetts 02138, USA}

 
\author{Bo Peng}
\email{bp432@cam.ac.uk}
\affiliation{Theory of Condensed Matter Group, Cavendish Laboratory, University of Cambridge, Cambridge CB3 0HE, United Kingdom} 

\author{Michele Pizzochero}
\email{mp2834@bath.ac.uk}
\affiliation{Department of Physics, University of Bath, Bath BA2 7AY, United Kingdom}
\affiliation{School of Engineering and Applied Sciences, Harvard University, Cambridge, Massachusetts 02138, United States}

\date{\today}

\begin{abstract}
We design antiferromagnetic spin-1/2 chains in fullerene nanoribbons by introducing extra C$_{60}$ cages at one of their edges. The resulting odd number of intermolecular bonds induces an unpaired $\pi$-electron and hence a quantised magnetic moment in otherwise non-magnetic nanoribbons. We further reveal the formation of an antiferromagnetic ground state upon the linear arrangement of spin-$1/2$ C$_{60}$ cages that is insensitive to the specific structural motifs. Compared with graphene nanoribbons, Janus fullerene nanoribbons may offer an experimentally more accessible route to magnetic edge states with atomic precision in low-dimensional carbon nanostructures, possibly serving as a versatile nanoarchitecture for scalable spin-based devices and the exploration of many-body quantum phases.
\end{abstract}

\maketitle


\section{Introduction}

Spin-1/2 chains are prototypical many-body systems featuring an inherently quantum mechanical spin degree of freedom where quantum criticality can be continuously tuned via the application of a magnetic field\,\cite{Motoyama1996,Hammar1999}. Theoretical models, most notably the Heisenberg antiferromagnetic chain, have long predicted exotic quantum properties\,\cite{Griffiths1964,Bonner1964,Mueller1981,Karbach1997}. Yet, only recent experimental advances have enabled their direct observation\,\cite{Sun2025,Su2025,Fu2025}, further motivating the search of one-dimensional materials as a platform to explore quantum magnetism and  correlated phenomena. In this vein, nanoribbons of graphene hosting $\pi$-electron magnetism\,\cite{Peng2025b,Song2025} have emerged as promising candidates, especially in light of their potential applications ranging from electronics\,\cite{Wang2021c}, spintronics\,\cite{Pesin2012,Slota2018}, or qubits\,\cite{Trauzettel2007,Guo2009,Recher2010,Droth2013}, to the realisation of topological \,\cite{Cao2017a,Groning2018,Sugimoto2020,Jiang2021a,Rizzo2021} and Majorana states\,\cite{Rizzo2018}. Nevertheless, achieving atomically precise, chemically stable graphene nanoribbons with well-defined magnetic edges, as well as reliably probing their magnetism, remains a significant challenge.

The recent synthesis of monolayer networks of C$_{60}$\,\cite{Hou2022} provides new opportunities for designing nanoribbons with tuneable edge geometries\,\cite{Peng2025c,Peng2025a} upon confinement in one dimension. Various nanoribbons can be envisioned from the different crystalline phases of fullerene networks, as predicted computationally\,\cite{Peng2022c,Peng2023,Jones2023,Wu2025,Shearsby2025,Kayley2025,Shaikh2025,Tromer2022,Ribeiro2022,Ying2023} and verified experimentally\,\cite{Meirzadeh2023,Wang2023,Zhang2025}. While the Kekul{\'e} valence structure of C$_{60}$ molecule has been well understood\,\cite{Klein1986,Austin1994,Rogers2001}, the possibility to realise $\pi$-electron magnetism in their extended networks by creating unpaired electrons remains unexplored. This approach could provide valuable insights for designing fullerene nanoribbon–based spin chains in the context of quantum magnetism.



Here, we use first-principles calculations to establish general design principles to achieve tailored magnetism in otherwise non-magnetic fullerene nanoribbons. This is accomplished by the introduction of extra C$_{60}$ cages at one of the edges, leading to the formation of Janus fullerene nanoribbons. As a result of the odd number of intermolecular bonds induced by the extra cage, one unpaired electron is created and confined within the C$_{60}$ molecule, acting as a spin-1/2 center. Neighbouring  C$_{60}$ units at the edges couple antiferromagnetically, effectively realising a quantum spin chain embedded in a fullerene nanoribbon. The magnetic edge states are robust for different
spatial arragenement of the extra cages along the edges. Overall, our findings enable the rational design and engineering of magnetic fullerene nanoribbons through a precise functionalisation of their edges.

\begin{figure*}
\centering
\includegraphics[width=\linewidth]{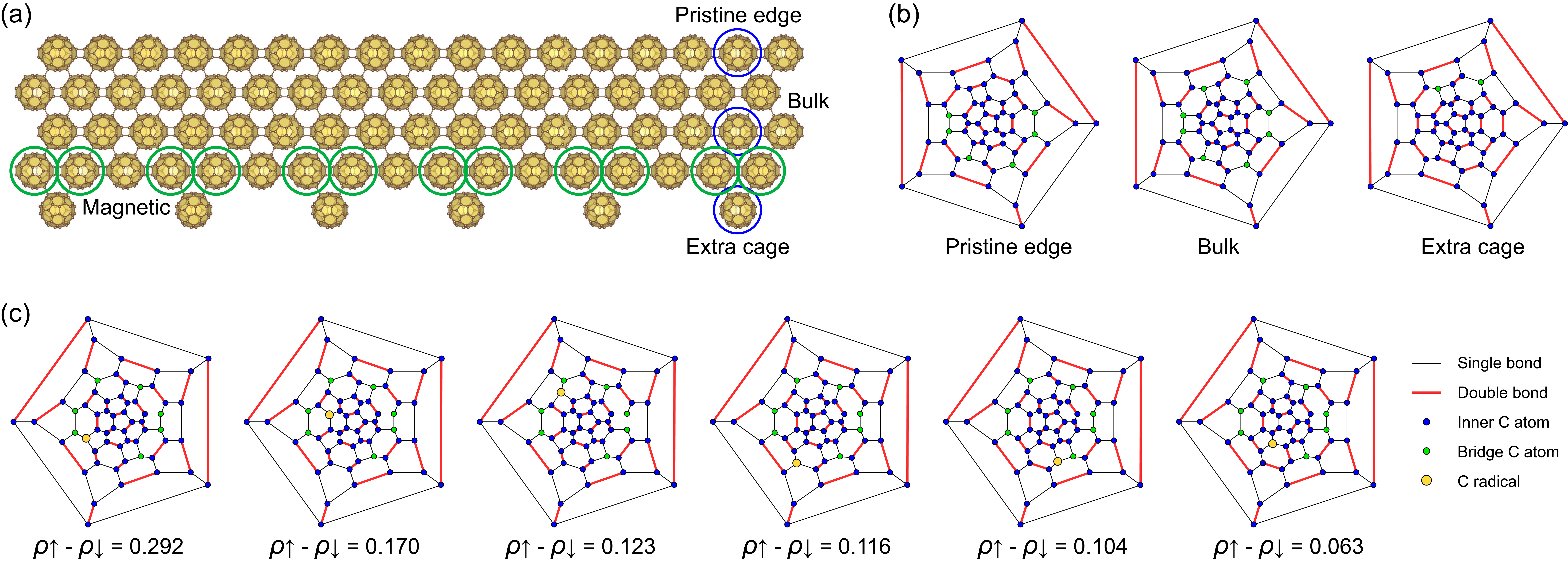}
\caption{
(a) Crystal structure of a Janus fullerene nanoribbons, along with the corresponding Schlegel diagrams for (b) non-magnetic and (c) magnetic C$_{60}$ cages.
}
\label{crystals} 
\end{figure*}

\section{Results and Discussion}

\subsection{Crystal structure} To introduce magnetic edge states, we add an extra C$_{60}$ cage at one of the edges  nanoribbon. This leads to the Janus fullerene nanoribbon shown in Fig.\,\ref{crystals}(a). Importantly, the addition of an extra C$_{60}$ cage to the pristine edge stabilises the system by 0.28\,eV, suggesting that Janus fullerene nanoribbons are likely to spontaneously form under thermodynamic equilibrium.
In Fig.\,\ref{crystals}(a), the C$_{60}$ cages denoted in green circles near the extra fullerene cage have an odd number of intermolecular bonds and hence host an unpaired electron, leading to $\pi$-electron magnetism. 

\subsection{Magnetic moments} To better understand the origin of $\pi$-electron magnetism in Janus fullerene nanoribbons, it is instructive to first consider non-magnetic, pristine fullerene nanoribbons and its constituent fullerene units. The resonance structure of each isolated C$_{60}$ cage features thirty double bonds and sixty single bonds. At the pristine edge of the nanoribbon displayed in Fig.\,\ref{crystals}(a), there are two intermolecular [2\,+\,2] cycloaddition bonds along the nanoribbon length and two intermolecular C--C single bonds across the nanoribbon width. The Schlegel diagram depicted in Fig.\,\ref{crystals}(b) shows one resonance structure of the fullerene in the pristine edge. The six carbon atoms forming intermolecular $sp^3$ bonds are indicated in green, which are fully saturated. For the other carbon atoms indicated in blue, the $sp^2$ hybridisation results in three single $\sigma$ bonds (the black line), while two neighbouring $\pi$ electrons join one of the three single bonds to form a double bonds (the red line). For the C$_{60}$ cages on the pristine edge, each unsaturated carbon atoms in blue is surrounded by two single bonds and one double bonds in Fig.\,\ref{crystals}(b), leaving no unpaired electron and hence no magnetic moment in the entire C$_{60}$ cage. Similarly, the fullerene cage in the bulk has no unpaired electron and is non-magnetic, as well as the extra fullerene cage, which is due to the even number of intermolecular bonds, as shown in Fig.\,\ref{crystals}(b). 

For C$_{60}$ cages in the green circles of Fig.\,\ref{crystals}(a), however, there are two intermolecular [2\,+\,2] cycloaddition bonds and three intermolecular C--C single bonds, leaving one unpaired electron. 
The unpaired electron in each magnetic C$_{60}$ cage with an odd number of interfullerene bonds has fully quantised spin-1/2, which is confined within the fullerene unit but delocalised at six main carbon sites due to the resonance structure. Fig.\,\ref{crystals}(c) shows the Schlegel diagram of the six magnetic carbon atoms in their resonance structure, confirming the quasi-localised nature of the unpaired electron (the yellow point). The Mulliken population difference between spin up and down ($\rho_{\uparrow}-\rho_{\downarrow}$) shows that these six carbon atoms contribute to more than 86.8\% of the total magnetic moment within the buckyball, i.e., 1\,$\mu_{\textrm{B}}$ per magnetic C$_{60}$.

\begin{figure*}
\centering
\includegraphics[width=\linewidth]{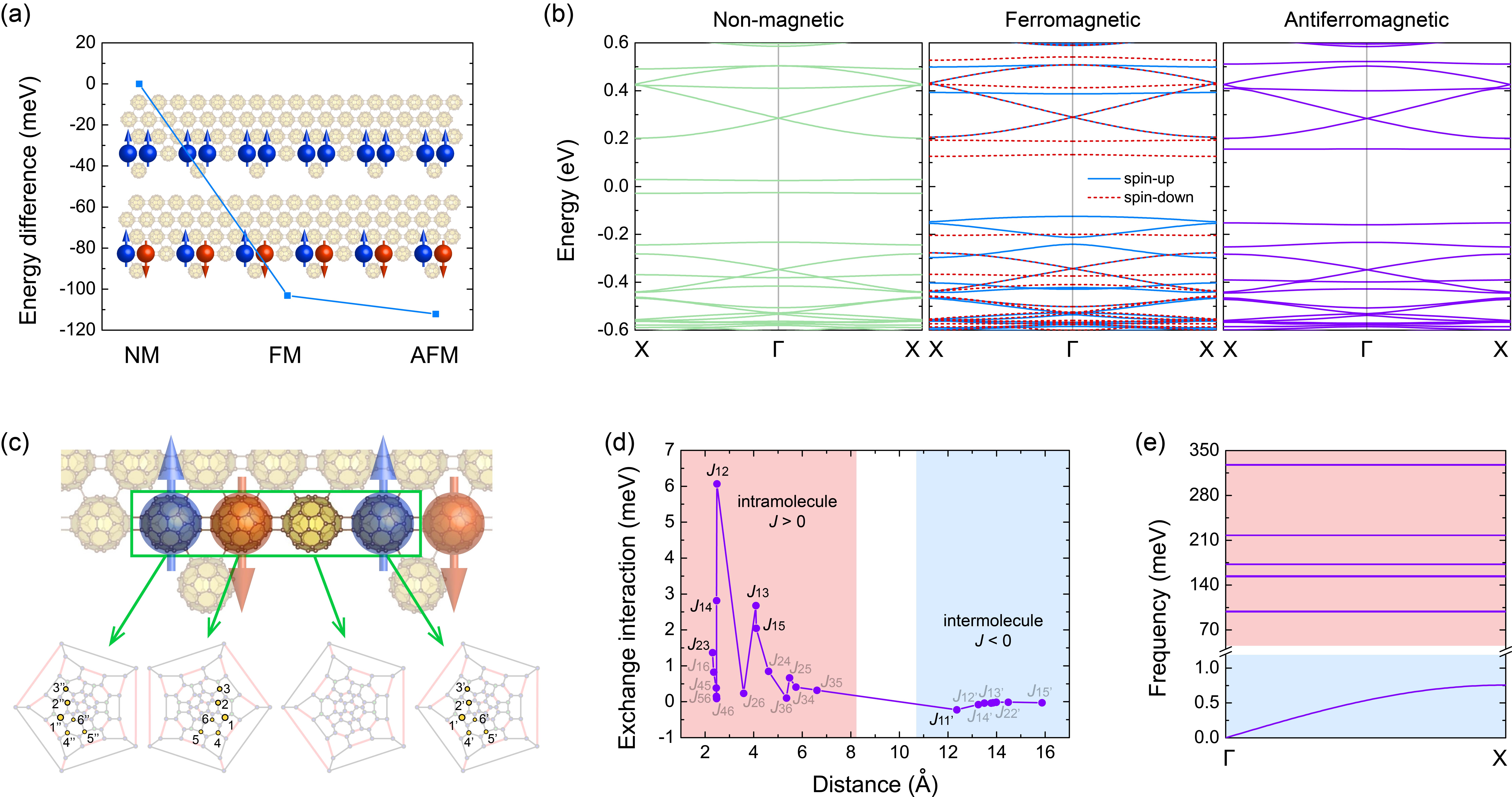}
\caption{
(a) Energy difference and (b) electronic band structure of Janus fullerene nanoribbons in the non-magnetic, ferromagnetic, and antiferromagnetic state. (c) Schlegel diagrams for antiferromagnetic fullerene chains. (d) Magnetic exchange interaction between magnetic carbon sites as a function of their distance, and (e) magnon spectrum.
}
\label{band} 
\end{figure*}

\subsection{Magnetic order} We next examine the magnetic order of the Janus fullerene nanoribbons. In Fig.\,\ref{band}(a), we  compare the difference in total energy between the non-magnetic, ferromagnetic, and antiferromagnetic phases, and also shown their corresponding spin orientations. The ferromagnetic phase is energetically more favourable by 103\,meV, while the antiferromagnetic nanoribbon is 9\,meV lower in energy. We thus conclude the antiferromagnetic is the most thermodynamically stable phase in the absence of external fields.

\subsection{Electronic  structure} We gain insights into the electronic properties of the three magnetic phases of Janus fullerene nanoribbons by studying their band structures, shown in Fig.\,\ref{band}(b). The non-magnetic phase has two flat edge states as the highest valence band and lowest conduction band respectively, with a band gap of 0.09\,eV. These two flat bands are mainly contributed by the fullerene units in green circles of Fig.\,\ref{crystals}(a), which act as defect-like states. In the ferromagnetic phase, while most of the bands remain doubly degenerate, the two top valence bands are spin-up states from the two magnetic fullerene cages, while the two bottom conduction bands are spin-down states from the same two C$_{60}$ cages. There are also several degenerate bands split into one spin-up band and one spin-down band around $-0.25$, 0.4, and 0.5\,eV respectively. For the stable antiferromagnetic phase, the deep valence bands and higher conduction bands remain the same with the non-magnetic phase, whereas the highest valence band and lowest conduction band exhibit a larger splitting with a band gap of 0.31\,eV. 


\subsection{Exchange interactions} We then examine the isotropic exchange interactions between magnetic carbon atoms as a function of the distance between these magnetic carbon sites. We label the magnetic sites in the red spin-down C$_{60}$ cages in green rectangle of Fig.\,\ref{band}(c) as $1-6$ from the largest to smallest magnetic moment. Within a C$_{60}$ cage ($<8$\,\AA), all intramolecular exchange interactions are ferromagnetic, as shown in Fig.\,\ref{band}(d). Only $J_{23}$, $J_{14}$, $J_{12}$, $J_{13}$, and $J_{15}$ have relatively larger exchange interaction ($>1$\,meV), especially $J_{12} = 6.065$\,meV. On the other hand, all intermolecular interactions are antiferromagnetic. Interestingly, the inter-fullerene interactions between the nearest neighbouring fullerene units, i.e., between $1-6$ and $1''-6''$ in Fig.\,\ref{band}(c), are all zero. However, the second nearest neighbouring C$_{60}$ cages, i.e., $1-6$ and $1'-6'$ in Fig.\,\ref{band}(c), have much stronger inter-fullerene exchange interactions, with the strongest one being $J_{11'} = -0.222$\,meV while the rest $|J_{ii'}| < 0.1$\,meV. This indicates weak antiferromagnetic interactions in the spin-1/2 chain. 

\subsection{Magnon spectrum} Using the calculated isotropic exchange terms, we obtain the magnon dispersion of the stable antiferromagnetic phase under the Holstein-Primakoff transformation\,\cite{Holstein1940} by diagonalising the bosonic Hamiltonian\,\cite{Colpa1978}. As shown in Fig.\,\ref{band}(e), the low-frequency magnons below 1\,meV are doubly degenerate and show linear dispersion around the $\Gamma$ point, indicating typical antiferromagnetic features. The magnon dispersion has non-negative values with a global minimum at $\Gamma$, demonstrating that the antiferromagnetic order is indeed the magnetic ground state\,\cite{Tellez-Mora2024}, in line with our energetic analysis.

\begin{figure*}
\centering
\includegraphics[width=0.8\linewidth]{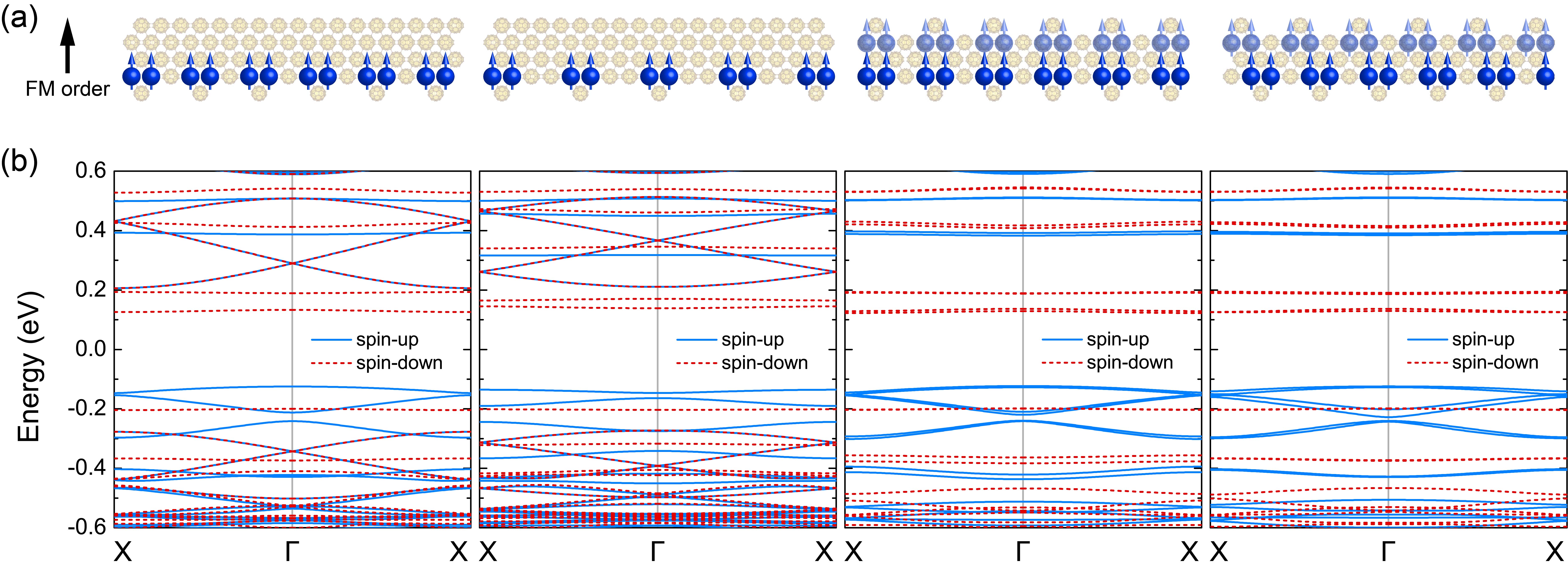}
\caption{
(a) Crystal structure of fullerene nanoribbons featuring various arrangements of extra C$_{60}$ cages at the edges and (b) their electronic band structure in the ferromagnetic phase.
}
\label{FM} 
\end{figure*}

\subsection{Robust magnetic edge states} Finally, we study the magnetic edge states for various spatial arrangements of extra C$_{60}$ cages at the edges. We focus on the ferromagnetic phase because of the spin-polarised nature of its magnetic edge states, that is particularly appealing for spintronics. On one hand, increase in the spacing between the extra C$_{60}$ cages located at the same edge, as shown in Fig.\,\ref{FM}(a), leads to an energy lowering of 0.63\,eV per C$_{60}$ unit. This agrees with our previous results showing enhanced thermodynamic stability of fullerene nanoribbons with increased number of fullerene cages in the unit cell\,\cite{Peng2025c}. On the other hand, nanoribbons with larger spacing between the extra C$_{60}$ cages exhibit distinct band structures for the ferromagnetic phase displayed Fig.\,\ref{FM}(b) but the same spin-1/2 bahaviour. Similarly, we can create chevron-like fullerene nanoribbons\,\cite{Ma2025} where both edges host $\pi$-electron magnetism doubling the magnetic moment per unit cell compared to Janus fullerene nanoribbons. For these chevron-like nanoribbons, two configurations are possible, depending on whether the two extra C$_{60}$ units on the opposite edges are aligned (leading to a space group of $P2/m$, No.\,10) or misaligned (space group $P\bar{1}$, No.\,2) across the nanoribbons. Despite the different space groups, however, the two ferromagnetic chevron nanoribbons have nearly identical band structures with a tiny total energy difference within 3.4\,meV.

\subsection{Experimental feasibility} The experimental realisation of quantum spin chains in fullerene nanoribbons is particularly promising due to several intrinsic advantages over other nanoribbons. Unlike graphene nanoribbons that require atomic precision at the edge and suffer from chemical instability, the fullerene-based nanoribbons are inherently robust, with chemically well-defined C$_{60}$ units and highly controllable intermolecular bonding configurations. Recent advances in the synthesis of monolayer fullerene networks\,\cite{Hou2022} and bottom-up arrangement of C$_{60}$ molecules with a scanning tunneling microscope tip\,\cite{Resh1994,Maruno1995} provide feasible routes to engineer the intermolecular bonding motifs that host $\pi$-electron magnetism. 

\subsection{Outlook} Our findings envision Janus fullerene nanoribbons as a versatile platform to design quantum spin chains for a wide range of applications from qubit entanglement to spintronics. Because of the intrinsically weak spin–orbit coupling and negligible hyperfine interaction in carbon\,\cite{Droth2013,Slota2018}, the spin coherence time is expected to be very long, making the proposed Janus fullerene nanoribbons promising building blocks for scalable qubit systems. Furthermore, individual spin-1/2 C$_{60}$ cages can serve as localised quantum dots with controllable exchange interactions. 
In addition, quantum chains realised by Janus fullerene nanoribbons can enable scalable architectures for quantum simulators, spintronic devices for exploiting spin-polarised currents, and platforms for exploring emergent topological phases and Majorana modes when coupled with superconductors. The same building blocks can also be extended beyond quasi-1D nanoribbons for realising even more complex condensed matter systems in 2D\,\cite{Wu2025a} such as ferromagnetic Haldane models\,\cite{Pingen2025} and altermagnetic Shastry-Sutherland lattice\,\cite{Wu2025b}. Overall, the combination of quantised spin, structural flexibility, and long spin coherence places Janus fullerene nanoribbons at the forefront of candidates for next-generation carbon-based quantum technologies. 

\section{Conclusions} 

In summary, our work establishes a strategy to create metal-free magnetism and one-dimensional spin chains based on 
Janus fullerene nanoribbons. By introducing extra C$_{60}$ cages to pristine edges, we demonstrate that $\pi$-electron magnetism can arise as a result of the odd number of intermolecular bonds, with quantised magnetic moments residing primarily at the C$_{60}$ cages. This gives rise to an antiferromagnetic $S=1/2$ chain with weak intermolecular exchange interactions. As compared to graphene-based systems, fullerene nanoribbons offer greater chemical stability and easier structural control, lowering the barrier for experimental realisation. Beyond their fundamental importance for probing quantum magnetism, these nanoribbons present promising opportunities for quantum technologies such as spintronic devices and qubit systems in virtue of the weak spin-orbit and hyperfine interactions intrinsic to carbon, which ensures long spin coherence times. Hence, the tuneability and scalability of these fullerene-based platform may pave the way toward integrating carbon-based magnetic architectures into next-generation quantum information systems.










\section*{Methods}

Density functional theory (DFT) calculations\,\cite{Hohenberg1964,Kohn1965} are performed under the spin-polarised, generalised-gradient approximation (GGA) of Perdew, Burke, and Ernzerhof (PBE) \cite{Perdew1996}, as implemented in the \textsc{SIESTA} package\,\cite{Soler2002,Artacho2008,Garcia2020}. A double-$\zeta$ plus polarisation (DZP) basis set is used with an energy cutoff of 400\,Ry and a reciprocal space sampling of 10 $k$-points along the periodic direction for structural relaxation. A vacuum spacing in the non-periodic directions larger than 20\,\AA\ is adopted throughout all the calculations. Both the lattice constant and atomic positions are fully relaxed using the conjugate gradient method\,\cite{Payne1992} with a tolerance on forces of 0.02\,eV/\AA. Our computational model of fullerene nanoribbons contain 660--1020 carbon atoms in the unit cell. For band structure calculations, the $k$-points are increased to 100 to sample the high-symmetry line. 
 For monolayer qHP networks, the inclusion of the Grimme's D3 dispersion corrections\,\cite{Grimme2010} leads to a decrease in lattice constants by merely 0.3\%\,\cite{Peng2022c,Shearsby2025}. We therefore neglect the van der Waals interactions hereafter. 
 For electronic structures, we use 100 $k$-points to sample the high-symmetry line. 
 
 The Mulliken population analysis between spin up ($\rho_{\uparrow}$) and down ($\rho_{\downarrow}$) is applied to study the magnetic moment at each carbon atom. We choose magnetic atoms with $|\rho_{\uparrow}-\rho_{\downarrow}|>0.06$, which contribute to nearly 90\% of the total magnetic moment. The exchange interactions between magnetic atoms are computed from the Green's function\,\cite{Liechtenstein1987,Korotin2015} based on the Wannier tight-binding Hamiltonian\,\cite{Marzari1997,Souza2001,Mostofi2008,Marzari2012,Mostofi2014,Pizzi2020}, as implemented in the \textsc{TB2J} package\,\cite{He2021}. Such interactions are found to vanish beyond 17\,\AA. For antiferromagnetic nanoribbons, the magnons are diagonalised based on the bosonic Hamiltonian\,\cite{Colpa1978} under the Holstein-Primakoff transformation\,\cite{Holstein1940}, as implemented in the \textsc{MAGNOPY} package that has been widely employed to study low-dimensional antiferromagnets\,\cite{Rybakov2024,Boix-Constant2025}. 




\section*{Acknowledgements}

We thank Dr Xu He at the University of Li{\`e}ge for helpful discussions on computing magnetic interaction parameters. B.P. acknowledges support from Magdalene College Cambridge for a Nevile Research Fellowship. The calculations were performed using resources provided by the Cambridge Service for Data Driven Discovery (CSD3) operated by the University of Cambridge Research Computing Service (\url{www.csd3.cam.ac.uk}), provided by Dell EMC and Intel using Tier-2 funding from the Engineering and Physical Sciences Research Council (capital grant EP/T022159/1), and DiRAC funding from the Science and Technology Facilities Council (\url{http://www.dirac.ac.uk} ), as well as with computational support from the UK Materials and Molecular Modelling Hub, which is partially funded by EPSRC (EP/T022213/1, EP/W032260/1 and EP/P020194/1), for which access was obtained via the UKCP consortium and funded by EPSRC grant ref EP/P022561/1.

\bibliography{references}

\end{document}